\documentclass[aps,prd,superscriptaddress,showkeys,showpacs,twocolumn,floatfix]{revtex4}

\usepackage{graphicx}
\newcommand{\beq}  {\begin{equation}}
\newcommand{\eeq}  {\end{equation}}
\newcommand{\bmath}{\begin{eqnarray}}
\newcommand{\emath}{\end{eqnarray}}

\def\lapproxeq{\lower .7ex\hbox{$\;\stackrel{\textstyle<}{\sim}\;$}}
\def\gapproxeq{\lower .7ex\hbox{$\;\stackrel{\textstyle>}{\sim}\;$}}
\def\be{\begin{equation}}
\def\ee{\end{equation}}

\begin{document}
\title{Superconvergence Relations and Parity Violating analogue of GDH sum rule.}
\author{Krzysztof Kurek}
\email{kurek@fuw.edu.pl}
\author{Leszek {\L}ukaszuk}
\email{lukaszuk@fuw.edu.pl}
\affiliation{Andrzej Soltan Institute for Nuclear Studies, Hoza 69, 00-681 Warsaw,
Poland}

\begin{abstract}
Sum rules of superconvergence type for parity violating amplitudes (p.v. analogue of 
Gerasimov-Drell-Hearn sum rule) are considered. 
Elementary processes initiated by polarized photons in the lowest order of 
electroweak theory are calculated as examples illustrating the validity of the p.v.sum
rules. 
The parity violating polarized photon-induced processes for proton
target are considered in the frame of effective low energy theories and phenomenological 
models based on p.v. nucleon-meson effective interactions.
Assuming the saturation of p.v. sum rule the possibility to limit
the range of the parameters, poorly known from 
existing experimental data and used in these models is discussed. 
The asymmetries for p.v. $\pi^0$ and $\pi^+$ production, measurable in future
high intensity polarized photon beams experiments, are given.

\end{abstract}

\maketitle

\section{Introduction}
The GDH sum rule~\cite{gdh} is recently 
intensively measured~\cite{rec1,rec2} and considered as a clean and important test of 
the knowledge of the nucleon spin structure especially in the resonance region ~\cite{dre}. 
The rising interest in GDH (and its $Q^2$ dependence generalization) 
and more generally in spin structure of nucleon has started with the new generation of
precise spin experiments~\cite{old0,old}.  
First direct data for real photons taken at MAMI~\cite{rec2}are especially 
important in understanding the subject and new data also with higher 
energies are now available and expected in future from ELSA, GRAAL, Jlab and
Spring-8~\cite{elsa,jlab,spring}.\\
The experiments based on intense polarized beams of photons~\cite{rec2,jp} 
give also the opportunity to test the weak (parity violating) part of photon-hadron 
interactions. The knowledge of 
p.v. couplings in nucleon-meson and nucleon-nucleon forces is a very important
point for understanding physics of nonleptonic weak p.v. hadronic interactions.
In addition the $\gamma\rho\pi$ and $\gamma\Delta N$  
p.v. couplings, very poorly known, can also play role  in photon-induced reactions.\\ 
It was shown in the paper~\cite{ji} that the polarized photon asymmetry
in $\pi^+$ photo-production near the threshold could be a good candidate to measure 
p.v. pion-nucleon coupling $h^1_{\pi}$. Similar expectations are connected with low energy
Compton scattering~\cite{beda,chen}.
Let us mention here that the $h^1_{\pi}$ coupling has been measured in nuclear~\cite{f} and atomic~\cite{c} systems.
The extraction of $h^1_{\pi}$ from such experiments is however difficult due to poor understanding 
of many-body systems. In fact, the disagreement between $^{18}F$ and $^{133}Cs$ experiments is seen~\cite{f,c}.

The experimental observation of p.v. effects in photonic reactions is generally difficult 
because the expected asymmetries are very small.
However, it seems sound to expect that the new high luminosity machines, generating intense polarized photon beams 
can change the situation in the nearest future~\cite{ji,jlablum}.  
Having this in mind a set of sum rules for parity violating part of Compton 
amplitudes has been recently derived by one of us (L.{\L})~\cite{ll}. 
In particular, p.v. analogue of GDH sum rule, based on Low Energy Theorems~\cite{let}
and under assumption of superconvergence of the type  
$\frac{f(z)}{z} \rightarrow 0$ at infinity for asymmetric amplitude was formulated there. 
In the past a number of superconvergence type sum rules for parity conserving Compton Scattering 
(one example of which reduces to GDH sum rule) has been obtained and discussed~\cite{ag,gdhdomb,letdombey}
and the superconvergence relations have been also studied  in detail in the very general 
context of axiomatic local field theory and its analyticity properties~\cite{mm}. 

In this paper we shall discuss p.v. analogue of GDH sum rule having essentially in mind two aspects:
verification and possible phenomenological implications.\\
The general formulae exploited in the paper are given in section 2. 
To verify p.v. sum rules for elementary targets we calculate in the lowest order of electroweak theory
the p.v. analogue of GDH sum rule integral for the 
processes with polarized photons scattered off leptons (section 3). 
The saturation hypothesis for p.v. sum rule  is discussed in section 4.
In section 5 the models of p.v. low energy photon-proton interactions (i.e.
heavy baryon chiral perturbation theory (HB$\chi$ PT)
~\cite{KM,chen} and low energy phenomenological models 
\cite{hen,des1,des2,des3}) are briefly described. 
Assuming the saturation for p.v. sum rule (similar to observed
quick saturation in GDH integral) we are able to narrow down allowed values of the
p.v. photon-meson and photon-$\Delta$-nucleon couplings (poorly known) and select the models 
with small high energy contribution (section 6). 
For these selected models 
the energy dependence of the asymmetries for pion photoproduction is calculated (section 7) according
to approach proposed in~\cite{hen}.
In the same section
it is shown that measurement of the photon energy dependence of the asymmetries from threshold up
to energy large enough to saturate p.v. sum rule (saturation energy) allows to distinguish between 
the different models which obey quick saturation feature. 
We conclude the paper with section 8.
For completeness the formulae of cross sections and asymmetries for p.v. photon-neutrino 
and photon-electron processes calculated in lowest order of electroweak theory
are given in Appendix A. 
The QCD perturbative cross sections formulae for unpolarized
photon scattered off polarized proton target are given in Appendix B.

\section{General formulae.} 
Let us consider p.v. Compton amplitudes for 
polarized photons scattered off unpolarized targets and for unpolarized photons off
polarized targets.
For the first case the following (crossing-antisymmetric) dispersion relation holds
(compare eq.(3.18) in~\cite{ll}):

\begin{equation}
Re f^{(-)\gamma}_h =\frac{\omega}{\pi} P 
\int_{\omega_{th}}^{\infty}\frac{
\omega'}{\omega'^2-\omega^2}(\sigma^T_h - \sigma^T_{-h})d\omega' + (subtr.),
\label{eq:1.1}
\end{equation}

where $\omega$ is the photon energy in laboratory system.
$f^{(-)\gamma}_h$ and  $\sigma^T_{h}$ are amplitude and relevant total cross 
section averaged over target particle spin, respectively; $h$ indicates photon helicity eigenvalue.

Second possibility brings us to the following (crossing-symmetric) formulae
(compare eq.(3.20) in~\cite{ll}):

\begin{equation}
Re f^{(-)tg}_s =\frac{1}{\pi} P \int_{\omega_{th}}^{\infty}\frac{
\omega'^2}{\omega'^2-\omega^2}(\sigma^T_s - \sigma^T_{-s})d\omega' + (subtr.). 
\label{eq:1.2}
\end{equation}

Here $f^{(-)tg}_s$ and $\sigma^T_s$ are amplitude and total cross section averaged over photon
polarization.\\

It was pointed out in~\cite{ll} that assumption of superconvergence for amplitude $f^{(-)\gamma}_h$ 
(i.e. no subtractions in~(\ref{eq:1.1})):
\begin{equation}
\frac{f^{(-)\gamma}_h(\omega)}{\omega}|_{\omega \rightarrow \infty} \rightarrow 0
\label{eq:1.3}
\end{equation} 
together with Low Energy Theorem (LET)~\cite{let} 
leads to 
the p.v.analogue of GDH sum rule (once limit $\omega \rightarrow 0$ is taken in unsubtracted~(\ref{eq:1.1})):

\begin{equation}
\int^{\infty}_{\omega_{th}}\frac{
\sigma^T_h - \sigma^T_{-h}}{\omega'}d\omega' = 0
\label{eq:1.4}
\end{equation}

For the polarized target case the relevant p.v. sum rule:

\begin{equation}
\int^{\infty}_{\omega_{th}}
(\sigma^T_s - \sigma^T_{-s})d\omega' = 0
\label{eq:1.5}
\end{equation}

can be obtained if stronger superconvergence is assumed:
~$f^{(-)tg}_s(\omega) \rightarrow 0$.

\section {The photon scattering off elementary lepton 
targets and verification of p.v. sum rules - illustrative examples.}
\begin{figure}
\begin{center}
\includegraphics[width=8cm]{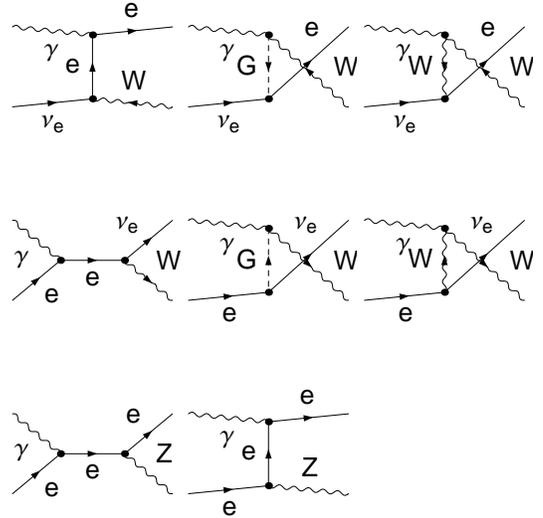}%
\end{center}
\caption{The Feynman  diagrams for lowest order p.v. processes induced by polarized photons} 
\label{fig:fdiag}
\end{figure}

As examples of the elementary parity violating processes three different inelastic 
polarized photon scattering off lepton targets are considered:\\ 
the photon-neutrino reaction into $W$ boson and electron, 
the photon-electron reactions into neutrino and $W$ boson and into 
electron and $Z^0$ boson production.
The relevant Feynmann diagrams in the lowest order of electroweak perturbation theory 
are depicted in 
fig.~\ref{fig:fdiag}. 

To be more general neutrino is treated as a massive particle in the photon-neutrino 
reaction. The sum rule~(\ref{eq:1.4}) can be rewritten as follows:

\begin{equation}
\int^{\infty}_{s_{th}}\frac{
\sigma^T_h - \sigma^T_{-h}}{(s-m^2)}ds = 0
\label{eq:2.1}
\end{equation}

where $s$ is energy square in CM system and $m$ is electron or neutrino mass.
In contrast to nonzero contribution for GDH integral in higher order of perturbation theory
the integrals in p.v. sum rules~(\ref{eq:1.4}) and~(\ref{eq:1.5})
should be zero at any order of perturbation theory. (More general discussion about p.c.sum rules 
and generalization of GDH sum rule including not only magnetic but also 
electron dipole moment can be found in~\cite{almo}.)\\
 The calculations have been done using 
{\it FeynArts 3} package for creating amplitudes and {\it HighEnergyPhysics} 
(HEP)~\cite{feyn} package for calculating relevant cross sections - both
 written for {\it Mathematica} system~\cite{volf}. 

The difference of the averaged cross sections over lepton helicity $\sigma^T_h - \sigma^T_{-h}$ in~(\ref{eq:2.1}) is 
expressed by the sum of $\frac{1}{2}(\Delta\sigma_{+} + \Delta\sigma_{-})$ where $\Delta\sigma_{\pm}$
are given in Appendix A, eq.~(9.4). It is straightforward to verify by elementary integrations that for every 
considered reaction (see formulae: (9.5) and (9.6) for $\gamma \nu \rightarrow We$ ,
(9.8) for $\gamma e \rightarrow Z^0 e$, (9.10) and (9.11) for $\gamma e \rightarrow \nu W$ ) the integral 
in eq.~(\ref{eq:2.1}) is zero for the difference of averaged lepton cross sections.

The reaction $\gamma \nu \rightarrow We$  with massless neutrino is simplest example 
of the violation of the second sum rule formulated in
section 2.in eq.~(\ref{eq:1.5}). Due to the fact that only one neutrino helicity state 
contributes to 
the cross section averaged over spin of photon (unpolarized photons case) the difference of 
the cross sections in the integral~(\ref{eq:1.5}) is always positive 
because it is simply averaged cross section of the reaction. 
Therefore there is no way to satisfy the condition to have
integral to be zero. The interesting question is whether this sum rule could be satisfied for hadron targets
when several different elementary processes might conspire to give vanishing difference 
of the total cross sections. We examine this possibility in the frame of perturbative QCD
approach and parton language structure of photon and proton target in Appendix B. 

\begin{figure}
\begin{center}
\includegraphics[width=8cm]{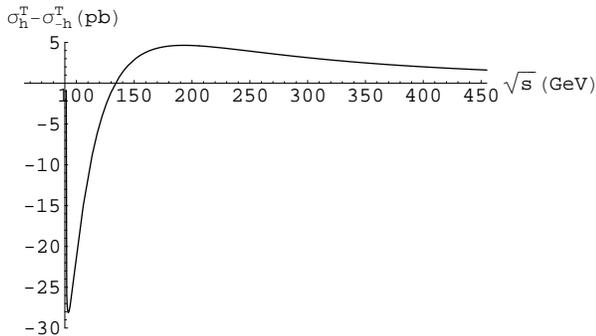}%
\end{center}
\caption{The difference of the polarized photon cross sections $\sigma^T_h - \sigma^T_{-h}$ for the reaction
$\gamma+e \rightarrow Z^0 +e$ as a function of energy.} 
\label{fig:zdsig}
\end{figure}

The typical behavior of the differences of cross sections in the sum rule 
~(\ref{eq:1.4}) is presented in fig.~\ref{fig:zdsig} for the reaction
$\gamma+e \rightarrow Z^0 +e$. It is seen in the fig.~\ref{fig:zdsig} that the saturation of the p.v. 
sum rule requires high energy contribution because the approaching of the differences of total cross 
sections to zero is slow.

Concluding this section let us add that
the photon-charged lepton reactions have been studied in the past in the frame of
electroweak theory. The contribution to GDH sum rule from the processes mediated by weak bosons 
in the lowest order of perturbation theory has been first time discussed in~\cite{altar} where also pure 
QED Compton scattering has been considered as well as Higgs boson production in this context.
Quite recently it was shown in \cite{dic} that GDH sum rule for electron evaluated at order of $\alpha^3$
agreed with the Schwinger contribution to the anomalous magnetic moment.

\section {The saturation hypothesis.}
One of the most interesting features of GDH sum rule for nucleon targets is
a quick saturation of the GDH integral. The dominant contribution (about 90\%) to the
GDH sum rule proceeds from the photon's energy range from the threshold up to $0.55$ GeV 
~\cite{rec1,rec2,elsa,jlab}. 
The saturation hypothesis in analogy with the feature
observed in the GDH sum rule is an important point in our analysis presented in section 6.
Therefore we are going now to formulate the criterion of the saturation of integral
in p.v. sum rule~(\ref{eq:1.4}). 
It is relatively easy to define the saturation when the integral in the sum rule has 
non zero value as it is in the case of GDH sum rule where the value of integral is determined by 
anomalous magnetic moment (with electric dipole moment 
if generalization of GDH is considered,~\cite{almo}) of target particle. 
The problem however appears when the integral in sum rule should be zero. 
Below we shall formulate the saturation criterion valid for both situations.
Given any superconvergence sum rule of the form:
\begin{equation}
a =  \int^{\infty}_{\omega_{th}}\frac{
\Delta \sigma(\omega')}{\omega'}d\omega',
\label{eq:3.4}
\end{equation}

we define the following $F$ quantity:

\begin{equation}
F(\omega) =  \frac{I_0}{I_1}.
\label{eq:3.1}
\end{equation}  
where:
\begin{equation}
I_0 = |a -  \int^{\omega}_{\omega_{th}}\frac{
\Delta \sigma(\omega')}{\omega'}d\omega'|,
\label{eq:3.2}
\end{equation}
and
\begin{equation}
I_1 = \int^{\omega}_{\omega_{th}}\frac{
|\Delta \sigma(\omega')|}{\omega'}d\omega',
\label{eq:3.3}
\end{equation}

Requirement that  $F(\omega)$ does not exceed prescribed small value at $\omega = \omega_{sat}$
determines saturation energy.
The usefulness of such definition of saturation is
based on the assumption that
there is no large contribution to the integral from photons with energy higher than 
$\omega_{sat}$.\\
For the GDH sum rule on proton, where experimental data~\cite{rec2,elsa} exist we can 
estimate $ \omega_{sat}$ to be  $0.55$ (i.e.$\frac{E_{sat}}{E_{th}} = 1.5$ in CMS) for $F(\omega_{sat}) = 0.1$.

As there are no experimental data for p.v. sum rules on proton we shall use the values $ \omega_{sat} = 0.6$ and
$F((\omega_{sat}) = 0.1$ in discussion of phenomenological consequences (sections 6 and 7).

\section{Proton target. The models of p.v. low energy photon-proton interactions. }

In this section we shall discuss two different approaches  to p.v. low energy photon-nucleon interactions.
We begin with  p.v. Compton amplitude on proton calculated in the frame of  HB$\chi$PT~\cite{beda,chen,KM}.
According to~\cite{chen} the  p.v. Compton amplitude can be written in CMS as follows:

\begin{eqnarray}
M^{(-)s_f,s_i}_{h_f,h_i}(\vec{k},\vec{k'}) &=&
\overline{N_{s_f}} [F_1\vec{\sigma}\cdot(\hat{\vec{k}} +\hat{\vec{k'}})
\vec{\epsilon_i}\cdot\overline{\vec{\epsilon'_f}}\nonumber\\
& -&
F_2 (\vec{\sigma}\cdot\overline{\vec{\epsilon'_f}} 
\hat{\vec{k'}}\cdot\vec{\epsilon_i}+ 
\vec{\sigma}\cdot\vec{\epsilon_i} 
\hat{\vec{k'}}\cdot\overline{\vec{\epsilon'_f}})\nonumber \\
&-&F_3 \hat{\vec{k}}\cdot\overline{\vec{\epsilon'_f}}
 \hat{\vec{k'}}\cdot\vec{\epsilon_i}\vec{\sigma}\cdot(\hat{\vec{k}} +
\hat{\vec{k'}})\nonumber\\
& -& i F_4\vec{\epsilon_i}\times\overline{\vec{\epsilon'_f}} 
\cdot(\hat{\vec{k}} +\hat{\vec{k'}})] N_{s_i},
\label{eq:4.1}
\end{eqnarray}

so that
\begin{equation}
f^{(-)p}_{\frac{1}{2}} = 2 F_1,
\label{eq:4.21}
\end{equation}
\begin{equation}
f^{(-)\gamma}_{h=+1} = -2 F_4.
\label{eq:4.2}
\end{equation}

To discuss p.v. sum rule and 
the superconvergence relations the interesting quantity is $F_4$ according to eq.~(\ref{eq:4.2}). 
The calculations based on HB$\chi$PT analysis in NLO~\cite{chen}  provides value of the coefficient
$F_4$ as follows:

\begin{equation}
F_4^{NLO} = -\frac{e^2 g_{A}h^1_{\pi}\mu_n}{8\sqrt{2}\pi^2 M F_\pi}
(\omega -\frac{m_\pi^2}{\omega}arcsin^2(\frac{\omega}{m_\pi})). 
\label{eq:4.3}
\end{equation}

It is easy to check that at high energies the real part of
$\frac{F_4^{NLO}}{\omega}$ converges to constant so superconvergence condition~(\ref{eq:1.3}) is
violated. Therefore the p.v. sum rule~(\ref{eq:1.4}) does not hold.

The similar formula with 6 independent structure functions $A_i$ can be written for p.c. 
Compton amplitude ~\cite{KM,chen}.
For the p.c. Compton amplitude the HB$\chi$PT gives the similar results for $A_3$ forward
scattering amplitude.
 According \cite{KM} $A_3$ is equal to:

\begin{equation}
A_3^{NLO}|_{\Theta=0} = -\frac{e^2 \omega \kappa_p^2}{2 M^2} -
\frac{e^2 g_{A}^2}{8\pi^2 F_\pi^2}
(\omega -\frac{m_\pi^2}{\omega}arcsin^2(\frac{\omega}{m_\pi})) 
\label{eq:4.4}
\end{equation}

which violates superconvergence relation of the type~(\ref{eq:1.3}) as in p.v. case. 

Discussing HB$\chi$PT it is important to note the fact that the spin-dependent p.c. 
polarizability  $\gamma_{p,n}$ (expressed by the integral similar to GDH integral but with 
higher power of energy in the denominator of integrand) essentially depends on loop
corrections and the contribution from the $\Delta$ and the lowest order result differs not only
in value but also in sign from corrected result~\cite{KM,chir03}.
Therefore a priori it is not excluded that 
p.v. sum rule~(\ref{eq:1.4}) might be satisfied if more 
corrections are taken into account in the frame of HB$\chi$PT. To our knowledge there is
no any $\chi$PT based model for p.v. Compton amplitude which obeys the superconvergence 
relation~(\ref{eq:1.3}).\\

Having this fact in mind 
we will consider existing in the literature low energy phenomenological model of pion photoproduction 
based on so-called pole approximation and effective Lagrangian's~\cite{hen}
(compare also~\cite{dombeyread})
and~\cite{des1,des2,des3}.
The model discussed in~\cite{hen} is relevant in low
energy regime so we will limit ourselves to energy below 0.55 GeV. The upper bound of energy 
is high enough to discuss and apply the saturation hypothesis as it was said in the previous
section. 
The contribution from high energy region will be ignored for a moment, 
assuming that is unimportant. \\
The detailed description of the approach can be found in~\cite{hen}.
The asymmetries of the polarized photon cross sections for $\pi^+$ and $\pi^0$
production are expressed by 
the sum of the p.v. coupling constants multiplied by the relevant form factors (see figs.11-15 in~\cite{hen}). 
In our calculations
the $\rho NN$ couplings $(h^0_{\rho},h^1_{\rho},h^2_{\rho})$, $\omega NN$ couplings $(h^0_{\omega},
h^1_{\omega})$ and $\pi N \Delta$ coupling ($f_\Delta$) have to be taken into account.
For $\pi^+$ production the important contribution follows from p.v. $\pi NN$ coupling
($h^1_{\pi}$). In addition there are two extra contributions from $\Delta$ 
directly  coupled to photon and nucleon ( $\gamma\Delta N$ coupling - $\mu^*$ ) and from  
interaction between photon, pion and $\rho$ meson ($\gamma\rho\pi$ coupling - $h_E$ ).
The last two parameters are directly related to the p.v. photon-mesons and photon-$\Delta$-nucleon
interactions while
the previous ones are related to strong sector (p.v. meson-nucleon couplings).\\
The knowledge of the values of p.v. couplings is rather limited; 
practically only ranges of values are known from experimental data (for review of the situation 
see \cite{des2} and references therein). 
On the other hand the strong sector meson-nucleon
couplings can be calculated in different approaches and models reviewed in~\cite{des2};
we shall exploit them in the next section. 
Especially difficult situation is for p.v.$\gamma\Delta N$ coupling  $\mu^*$ 
and $\gamma\rho\pi$ ($h_E$) which are given in the models from~\cite{des2}.
Only quite large limits  $\mu^* \in(-15,15)$ and $h_E \in(-17,17)$ in units $10^{-7}$ 
can be given for these couplings based on data analysis.

The $\mu^*$ and $h_E$ couplings can be calculated if some extra assumptions are added.
The vector-meson dominance model have been used in~\cite{hen} to 
estimate the p.v. $\gamma\Delta$N coupling . Neglecting $\omega$ and $\phi$ meson's
contribution and assuming that p.v. $\rho\Delta$N coupling is by size similar to 
p.v. $\rho$NN coupling the following relation have been formulated~\cite{hen}:

\begin{equation}
\frac{\mu^*}{M}=\frac{h_\rho^0}{g_\rho m_\rho}.
\label{eq:4.5}
\end{equation}  

 Taking $g_\rho$ equal to $0.43 4\pi$ (after \cite{hen}) the $\mu^* \simeq 0.55 h_\rho^0$
is uniquely defined by $h_\rho^0$ coupling. 

The $h_E$ coupling can be calculated according analysis described in~\cite{her}. Assuming so-called 
factorization ~\cite{her} the following relation can be written for $h_E$:

\begin{equation}
h_E = -2 M G^{\rho\pi\gamma},
\label{eq:4.6}
\end{equation}  
and
\begin{eqnarray}
G^{\rho\pi\gamma}&=&\sqrt{4\pi\alpha} G^{\rho\pi} (1\pm \frac{m_{a_1}^2 m_\rho g_T^a}
{\sqrt{2}f_\rho (m_{a_1}^2 - m_\rho^2)}\nonumber\\
&\mp& \frac{\sqrt{2}}{20}\frac{m_\rho g_T^b}
{f_\omega(m_{b}^2 -m_\rho^2)}).
\label{eq:4.7}
\end{eqnarray}  

Taking into account the present data~\cite{data} of the widths of $a_1$ and $b$ resonances, their 
masses and that $G^{\rho\pi}$ is $8.9\cdot 10^{-7}$ the two possible solutions have been found for
$h_E$: $\sim 10\cdot 10^{-7}$ and $\sim 1\cdot 10^{-7}$. The results are close to the ones 
obtained in $SU(6)_W$ based calculations in~\cite{her}.

In the next section the p.v. sum rule~(\ref{eq:1.4}) together with saturation criterion
will be used  to select the models which posses quick saturation feature.

\section{Phenomenological consequences of saturation.}

The p.v. meson-nucleon coupling constants calculated from
the flavor-conserving part of the quark weak interactions are reviewed in~\cite{des2}. 
The eight sets of numerical predictions for the p.v. meson-nucleon coupling constants,
calculated with different assumptions and models have been summarized in Table~\ref{models},
taken from~\cite{des2}(Table 1).
\begin{table}[htb]
\begin{tabular}{l r r r r r r} \hline  \hline \\
Model & $h^1_{\pi}$ & $h^0_{\rho}$ & $h^1_{\rho}$ & $h^2_{\rho}$ & $h^0_{\omega}$ &$h^1_{\omega}$\\ 
\hline  \\
1. DDH, range (K=6);ref.\cite{des1}   & 0.0 & 11.4 & 0.0 & -7.6 & 5.7 &-1.9  \\
2. DDH, range (K=6);ref \cite{des1}   & 11.4 & -30.8 & 0.4 & -11.0 & -10.3 &-0.8  \\ \\
3. DDH, (``best'');ref \cite{des1}   & 4.6 & -11.4& -0.2 & -9.5 & -1.9 &-1.1  \\ \\
4. D, range (K=3);ref \cite{des3}   & 1.3& 8.3& 0.0 & -8.2 & -0.5 &-1.8  \\
5. D, range (K=3);ref \cite{des3}   &2.0&-23.1& -0.3 & -8.2 & -10.6 &-2.2  \\ \\
6. D, range (K=1);ref \cite{des3}   &0.5&7.0& -0.2 & -10.3 & 2.5 &-2.0  \\
7. D, range (K=1);ref \cite{des3}   &0.4&-29.5&0.0 & -10.3 & -10.2 &-1.1  \\ \\
8. KM ;\cite{KM}   &0.2&-3.7& -0.1 & -3.3 & -6.2&-1.0  \\  \\ \hline \hline
\end{tabular}
\caption{Predictions for the p.v. meson-nucleon coupling constants after~\cite{des2}}
\label{models}
\end{table}

The predictions of the p.v. couplings are grouped in
five groups depending on the method of calculations (as in ~\cite{des2}).
The strong effects are partially incorporated in the meson's and nucleon's wave 
functions and partially in bare quark interactions. This part of strong interaction 
corrections manifests  themselves via dependence on K parameter (K=1 in the absence of 
corrections) in effective quark interactions.
The first two groups called in our notations as models 1-3 (DDH), are
based on the calculations from~\cite{des1}. Models from first group (1-2) contain strong 
corrections characterized by K=6 range parameter.  
The third and fourth group 
(models 4-5 and 6-7, D) have been calculated in~\cite{des3}. 
The models 4 and 5 are corrected for strong interactions (K=3) while models
6 and 7 (K=1) have no strong corrections taken into account. 
The predictions for model 4 and 6 correspond to the factorization
approximation used in the calculations while 5 and 7 are the results obtained assuming the 
validity of the $SU(6)_W$ symmetry.
The last set of couplings (model 8, KM ) is based on HB$\chi$PT
calculations,~\cite{KM}.               
All couplings are in units $10^{-7}$ and the $h_{\rho}^{1`}$ coupling, 
(originally listed in Table 1. from~\cite{des2}) has been omitted;  in fact it is zero 
for all models except KM.  $h_{\rho}^{1`}$ is not used in our calculations as it does not enter
in Henley et al. approach~\cite{hen} .\\

The values of p.v. couplings listed in the table will be used to verify the quick 
saturation feature  according the approach discussed in previous section and in~\cite{hen}. 

The $\mu^*$ and $h_E$ couplings will be 
treated as free parameters in the range limited by the 
experimental knowledge: $\mu^* \in(-15,15)$ and $h_E \in(-17,17)$ in units $10^{-7}$.
The contribution related to $f_\Delta$ parameter
is very small in the considered approach; we have fixed this coupling to be $10^{-7}$ after~\cite{hen}. 

Taking the values of p.v. coupling constants from Table~\ref{models} the
differences of polarized photon cross sections have been calculated and used for 
estimation of $F$ defined in~\ref{eq:3.1}. The saturation expressed by the condition 
$F<0.1$ limits the allowed values of $\mu^*$ and $h_E$ couplings.

According to~\cite{hen} the most important contribution to the differences of the
cross sections and asymmetries comes from
$\pi NN$ interaction if $h^1_{\pi}$ is not very small. It is expected that strong
interaction corrections will increase the value of $h^1_{\pi}$~\cite{des2}.
Indeed it is seen in Table~\ref{models} for D models (group 3 and 4) that increasing 
range K the $h^1_{\pi}$ also increases (models 4 and 5 with K=3 versus models 6 and 7,
not corrected, K=1). On the other hand $h^1_{\pi}=0$ is also not excluded for models
with higher correction range K=6 (model 1 from DDH group). The small value of 
$h^1_{\pi}$ is also supported by calculations of~\cite{KM}(model 8).

Apart from the model 2 and ``best fit'' model 3 from DDH group, quick saturation can be achieved 
for all other models from Table~\ref{models} by limiting range of free parameters  $\mu^*$ and $h_E$.
The values
of $\mu^*$ and $h_E$ obtained in~(\ref{eq:4.5}) and in the analysis from~\cite{her} also allow
to have saturation for models 4-7; the model 8 (KM) with these values of $\mu^*$ and $h_E$ has no saturation.
Experimental consequences of models with saturation are given in section 7.\\

It should be emphasized that the non saturating models (2 and 3) characterize large values
of $h^1_{\pi}$.
In consequence  it is impossible to find any pair of $\mu^*$ and $h_E$ couplings in allowed
range to satisfy the saturation condition. This observation leads to the conclusion 
that for these models some additional 
structure should be observed for higher than $0.55$ GeV photon's energy (compare
discussion and eq.(4.14) in ref.~\cite{ll}).\\
Considering quick saturation as a universal feature,
related to the complexity of hadronic targets 
the absence of saturation can be treated as an argument against large value 
of $h^1_{\pi}$. \\

\section{Pion photoproduction asymmetries.} 
In this section we are going to discuss the experimental consequences of the 
models satisfying quick saturation i.e. the  $\pi^+$ and $\pi^0$ p.v. asymmetries.

\begin{figure}
\begin{center}
\includegraphics[width=8cm]{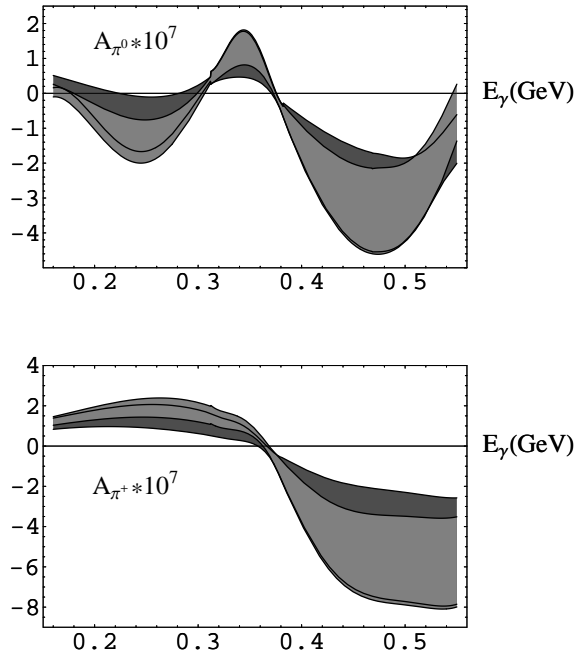}%
\end{center}
\caption{The asymmetries for $\pi^+$ and $\pi^0$ photoproduction as a function of photon energy.
The shadowed bands reflect freedom of the values of p.v. photon-meson couplings $\mu^*$ and  
$h_E$ allowed by saturation condition for model 4.
The darker band is for positive, lighter for negative values of $h_E$. $\mu^*$ 
is always negative in shadowed bands.}
\label{fig:asy4}
\end{figure}

\begin{figure}
\begin{center}
\includegraphics[width=8cm]{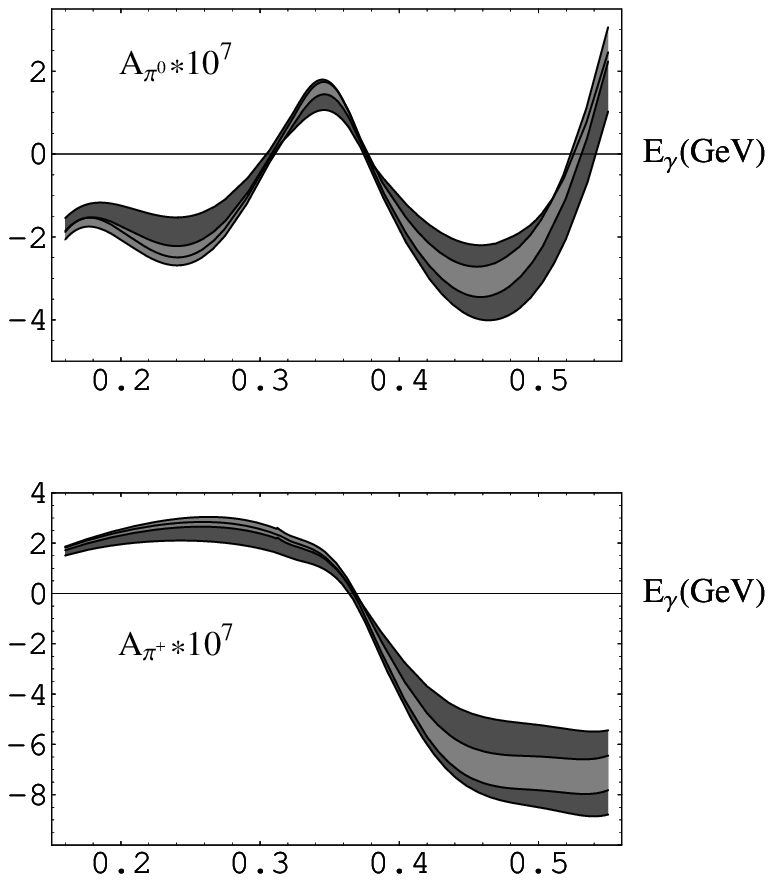}%
\end{center}
\caption{The asymmetries as in fig.~\ref{fig:asy4} for model 5. } 
\label{fig:asy5}
\end{figure}

\begin{figure}
\begin{center}
\includegraphics[width=8cm]{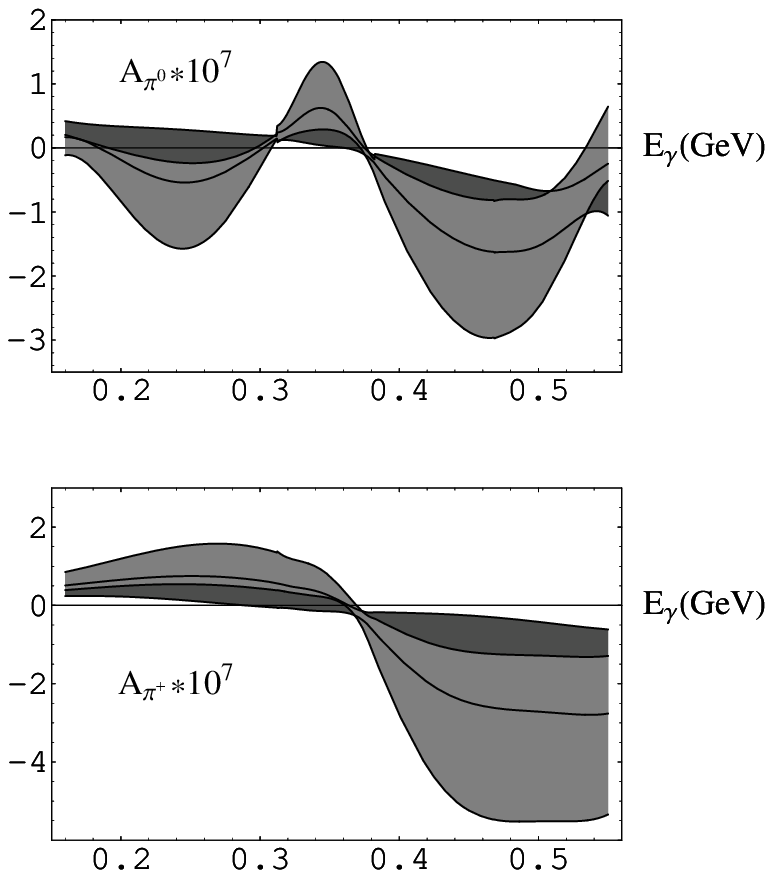}%
\end{center}
\caption{The asymmetries as in fig.~\ref{fig:asy4} for model 6. } 
\label{fig:asy6}
\end{figure}

\begin{figure}
\begin{center}
\includegraphics[width=8cm]{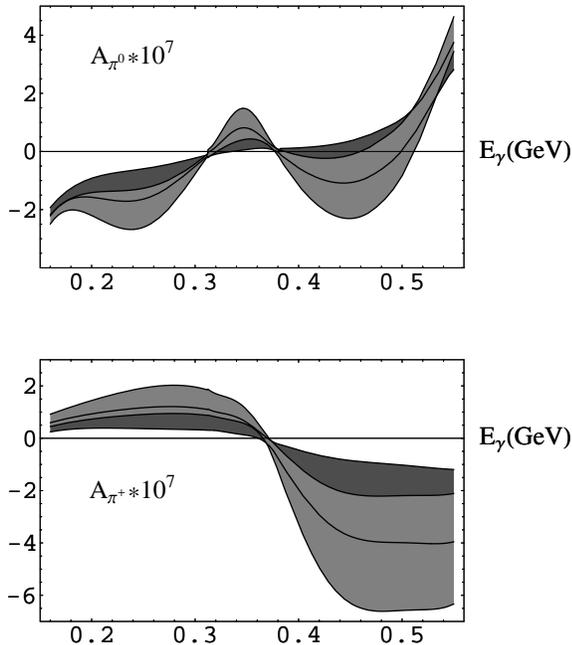}%
\end{center}
\caption{The asymmetries as in fig.~\ref{fig:asy4} for model 7. } 
\label{fig:asy7}
\end{figure}
 
In the fig.~\ref{fig:asy4} and~\ref{fig:asy5} the pion photoproduction asymmetries are
shown for strong interactions corrected models from D groups,model 4 and 5.
The $\pi^+$ asymmetries are positive at threshold,
relatively large and negative for photon energy close to $0.5$ GeV.
The $\pi^0$ asymmetries on threshold are sensitive to assumptions under which
the predictions for couplings have been calculated; the factorization (model 4) prefers
zero or very small and rather positive asymmetry  while  $SU(6)_W$ symmetry (model 5) 
leads to larger and negative asymmetries.

The models 6 and 7 from D groups in which the strong interaction corrections have not been
taken into account characterize smaller value of leading coupling $h^1_{\pi}$ and therefore 
the saturation is possible with much more freedom in photon-meson couplings than for
corrected models 4 and 5. The asymmetries are similar in shape however smaller and less
constrained (figs.~\ref{fig:asy6} and~\ref{fig:asy7}). 
 
\begin{figure}
\begin{center}
\includegraphics[width=8cm]{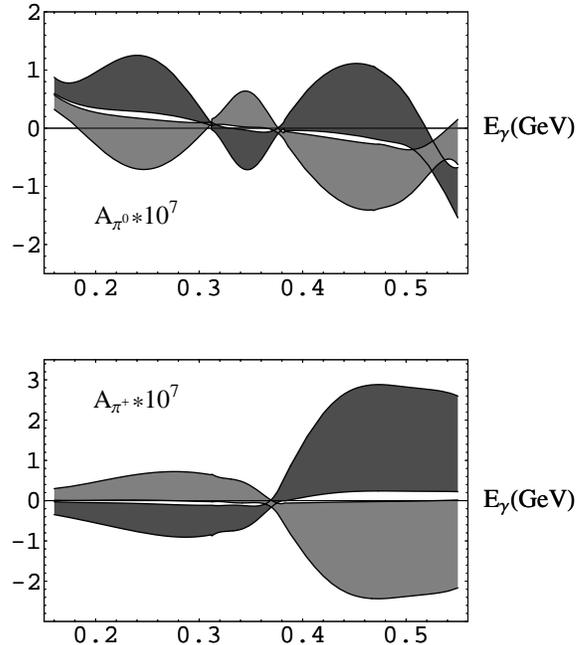}%
\end{center}
\caption{The asymmetries as in fig.~\ref{fig:asy4} for model 1. Here the darker band is for
positive values of $\mu^*$ and  $h_E$ couplings, lighter for negative.} 
\label{fig:asy1}
\end{figure}
 
Model 1 from DDH group and 8 (KM) are characterized by smallest couplings $h^1_{\pi}$
(0 in the case of model 1). The saturation is possible with large freedom in couplings 
$\mu^*$ and  $h_E$ (model 1,fig.~\ref{fig:asy1}).
The asymmetries for model 8 (KM) are similar to those presented
in fig.~\ref{fig:asy1}, reaching $-4*10^{-7}$ for higher energy. 
For the models with small or zero p.v. pion-nucleon $h^1_{\pi}$ the sign of $\mu^*$ and  $h_E$ 
couplings is related directly to the sign of $\pi^+$ asymmetry. \\

As a summary of this section we conclude that 
the most interesting models with quick saturation 
are models 4 and 5 from D group with corrections range K=3, based on calculations done 
in~\cite{des3}; the predicted values of pion photoproduction asymmetries are relatively large.
While the $\pi^+$ asymmetries are similar in energy dependent shape for different models,
$\pi^0$ asymmetries at threshold showed the interesting model dependence (compare
fig.s~\ref{fig:asy4}-~\ref{fig:asy7})

\section{Concluding remarks.}

We have verified p.v. superconvergent sum rules formulated in~\cite{ll}
and we have examined their phenomenological consequences;
the sum rules have been checked within the lowest perturbative order of 
electroweak theory for the photon induced processes with elementary lepton targets.
The p.v. analogue  of GHD sum rule for polarized photons interacting with unpolarized
targets have been verified by straightforward calculations.\\
It would be interesting to check this sum rule in higher perturbative orders as it was 
recently done for GDH in QED in~\cite{dic}.\\
In analogy with GDH sum rule for nucleon the saturation hypothesis has been
formulated and the eight models with different sets of p.v. couplings~\cite{des2}
have been analyzed using p.v. photoproduction model proposed in~\cite{hen}.
The models with largest leading p.v. pion-nucleon coupling $h^1_{\pi}$ do not saturate 
p.v. sum rule integral  below energy of photon smaller than $0.6 GeV$ and the contributions 
from higher energies cross sections are needed. 
It suggests some structure in difference of the cross sections to be observed 
for higher photon's energy for these models.\\
The other models considered in the paper  have enough freedom in parameter space 
defined by data and calculations to saturate p.v. sum rule integral below photon's energy 
$0.55$ GeV.\\
The p.v. asymmetries have been calculated in the photon energy range from
threshold to $0.55$ GeV. The asymmetries for some
models satisfying saturation are large enough to be measured in the future experiments with
intensive beams of polarized photons.

\section{Appendix A.}
Below the set of formulae describing polarized cross sections calculated in the lowest order
of electroweak perturbative theory is presented.
We consider the photon-neutrino scattering into $W$ boson and electron, the scattering of 
photon-electron into neutrino and $W$ boson and into electron and $Z^0$ boson.

Let us define the general formula for the cross sections as follows:
  
\begin{eqnarray}
\sigma_n(r_1,r_2,r_3)& = &\frac{\pi \alpha^2}{2 sin(\Theta_W)^2 M_W^2 (1-r_1)^n} 
[w_1^n P(r_2,r_3)\nonumber\\
 &+& w_2^n L(r_2,r_3) + w_3^n L(-r_2,r_3)] 
\label{eq:a.1} 
\nonumber
\end{eqnarray}

where: 
\begin{eqnarray}
P(x,y) &=& \sqrt{1 + x^2 + y^2 -2(x + y + x y)},\nonumber \\
L(x,y) &=& ln(\frac{1 + x + y + P(x,y)}{1 + x + y - P(x,y)}).
\label{eq:a.2} 
\end{eqnarray}
$r_i$ are ratios $\frac{m_i^2}{s}$ and $m_i$ are masses of particles taking part in reactions.
$w_i^n$ are coefficient functions depending on the reaction (see subsections below).

The unpolarized (averaged over photon and lepton helicity) cross sections $\sigma$ and 
fully polarized ones
$\sigma^h_s$ are equal to:
 
\begin{eqnarray}
\sigma &=&\frac{1}{4}( \sigma^h_s + \sigma^{-h}_s+ \sigma^h_{-s} + \sigma^{-h}_{-s})
\label{eq:a.3} 
\end{eqnarray}

Here $h$ refers to photon and $s$ to lepton helicity ($s,h = \pm$).

The differences of the cross sections: $\Delta\sigma$ and $ \Delta\sigma_s$
are defined as follows:

\begin{eqnarray}
\Delta\sigma &=&\frac{1}{2}( \sigma^h_s + \sigma^{-h}_s- \sigma^h_{-s} - \sigma^{-h}_{-s}),\nonumber \\
\Delta\sigma_s &=&\sigma^h_s - \sigma^{-h}_s. 
\label{eq:a.4} 
\end{eqnarray}

$\sigma$ and $\Delta\sigma$ have a form of $\sigma_3$; \\
$\Delta\sigma_s$ is $\sigma_2$ type.\\

The coefficient functions $w_i^3$ and $w_i^2$ are process-dependent.
We listed them in subsections.

\subsection{Cross sections for $\gamma \nu \rightarrow We$}

The cross sections are:
\begin{eqnarray}
\sigma &=&\sigma_3(r_{\nu},r_e,r_W), \nonumber\\
\Delta\sigma_{-}&=&\sigma_2(r_{\nu},r_e,r_W), \nonumber\\
\Delta\sigma_{+}&=&0 \nonumber\\
\Delta\sigma&=&-2 \sigma.
\label{eq:a.5} 
\end{eqnarray}

The coefficient functions $w_i$ are the following:
\begin{eqnarray}
w_1^2 &=&-4 (r_{\nu} - r_e + 6 r_W), \nonumber\\
w_2^2 &=& 2 (1 + r_{\nu})(r_{\nu} - r_e) + 4 r_W (1-2 r_W - 3 r_e), \nonumber\\
w_3^2 &=& 4 r_W (2 + r_{\nu} -2 r_W - 3 r_e), \nonumber\\
w_1^3 &=& 4 (1 - r_e^2 - r_e (r_W -2 r_{\nu}) + 2 r_W^2 - r_{\nu} (2+r_W)),\nonumber\\
w_2^3 &=& (1 - r_{\nu})^2 (r_{\nu} + r_e + 2 r_W),\nonumber \\
w_3^3 &=& 2 (1 - r_e -r_W)((r_e - r_{\nu})^2 + r_W (r_{\nu} + r_e -2 r_W)), \nonumber\\
r_{\nu} &=& \frac{m_{\nu}^2}{s},\nonumber\\
r_e &=& \frac{m_e^2}{s},\nonumber\\
r_W &=& \frac{m_W^2}{s}.
\label{eq:a.6} 
\end{eqnarray}

\subsection{Cross sections for $\gamma e \rightarrow Z^0 e$}

The cross section are:
\begin{eqnarray}
\sigma &=&\sigma_3(r_e,r_e,r_Z), \nonumber\\
w_1^3 &=& 2 r_Z ((1 -\beta) r_Z - (1 + 2 \beta) r_e) \nonumber\\
&+& 
\frac{1}{4}(1 - r_e)^2 ((1 - r_e)^2 - r_Z^2 - (1 +\beta r_Z)(1 + r_e - r_Z)),\nonumber\\
w_2^3 &=& \frac{1}{2} (1 - r_e)^2 + \frac{3}{2}(1 +\beta) r_Z (1 + r_e)^2 \nonumber \\
&-&
r_Z (1 + 2 \beta + 3 r_e) - r_Z^2 ((1 -\beta) (1 - r_Z) + r_e(2 + \beta)), \nonumber\\
w_3^3 &=& 0, \nonumber\\
r_Z &=& \frac{m_Z^2}{s}, \nonumber \\
\beta &=& 4 sin(\Theta_W)^2 (1 - 2 sin(\Theta_W)^2).
\label{eq:a.7} 
\end{eqnarray}
and
\begin{eqnarray}
\Delta\sigma_{\pm}&=&\sigma_2(r_e,r_e,r_Z), \nonumber\\
w_1^2(+) &=& \frac{1}{2}(- r_e^3+ r_e^2(6 +r_Z (\beta+\delta)) \nonumber\\
&+& r_e (7+r_Z^2 (1-\beta-\delta) - r_Z (1+ 2 \beta + 2 \delta)) \nonumber \\
&-& r_Z (5+3 r_Z)) (1-\beta -\delta)), \nonumber\\
w_2^2(+) &=& 5 r_e^2 + r_e (1 - r_Z (3- \beta-\delta)) \nonumber \\
&-& r_Z (1  -2 r_Z) (1 - \beta -\delta), \nonumber\\
w_3^2(+) &=& w_3^2(-) = 0, \nonumber\\
w_1^2(-) &=& \frac{1}{2}(r_e^3 - r_e^2(6 +r_Z (\beta - \delta)) \nonumber\\
&-& r_e (7+\frac{1}{2}r_Z^2 (1+\delta)^2 - r_Z (1+ 2 \beta - 2 \delta)) \nonumber \\
&+& \frac{1}{2}r_Z (5+3 r_Z) (1 + \delta)^2), \nonumber\\
w_2^2(-) &=& -5 r_e^2 - r_e (1 - r_Z (3- \beta + \delta)) \nonumber\\
&+& \frac{1}{2}r_Z (1  -2 r_Z) (1 +\delta)^2, \nonumber\\
\delta &=& 1 - 4 sin(\Theta_W)^2.
\label{eq:a.8} 
\end{eqnarray}

\begin{eqnarray}
\Delta\sigma&=&\sigma_3(r_e,r_e,r_Z), \nonumber\\
w_1^3 &=& -\frac{1}{2}\delta r_Z (1 + 7 r_Z + r_e^3  - r_e^2 (5 + r_Z)\nonumber \\
&-& 3 r_e (7 - 2 r_Z)), \nonumber\\
w_2^3 &=& \delta r_Z ((1  - r_Z^2) + r_Z^2 + r_e (2 + 9 r_e -6 r_Z)),\nonumber\\
w_3^3 &=& 0.
\label{eq:a.9}
\end{eqnarray}

\subsection{Cross sections for $\gamma e \rightarrow \nu W$}
The cross sections are:

\begin{eqnarray}
\sigma &=&\sigma_3(r_e,r_W,0), \nonumber\\
w_1^3 &=& \frac{1}{4}(r_e^3 (1 - r_W) - 2 r_e^2 (1 -6 r_W + r_W^2) \nonumber\\
&-& 
r_e (15 + 29 r_W -4 r_W^2) + 2 (4 + 5 r_W + 7 r_W^2)), \nonumber\\
w_2^3 &=& r_e^2 (1 - r_W) + r_W (r_e (9 + r_W) -2 (2 + r_W + r_W^2)), \nonumber \\
w_3^3 &=& 0, 
\label{eq:a.10} 
\end{eqnarray}
and

\begin{eqnarray}
\Delta\sigma^{\pm}&=&\sigma_2(r_e,r_Z,0), \nonumber\\
w_1^2(+) &=& r_e (1 + r_e  + r_W (3- r_e)),  \nonumber   \\
w_2^2(+) &=& - 4 r_W r_e,    \nonumber \\
w_3^2(+) &=& w_3^2(-) = 0, \nonumber \\
w_1^2(-) &=& - 2 r_W (13 + 3 r_W + r_e (5 -  r_W)),  \nonumber   \\
w_2^2(-) &=& 8 r_W (1 + r_e + 3 r_W), 
\label{eq:a.11} 
\end{eqnarray}

\begin{eqnarray}
\Delta\sigma &=& \sigma_3(r_e,r_W,0), \nonumber\\
w_1^3 &=& \frac{1}{2} r_e ((1 + r_e)^2 - r_W (r_e^2 + 2 r_e (2 - r_W) \nonumber\\
&-& 3 (1- 4 r_W))) - 4 (1 - 2 r_e) - r_W (5 + 7 r_W), \nonumber\\
w_2^3 &=& 2( - r_e^2 (1 - r_W) - r_e r_W (7 - 5 r_W)  \nonumber\\
&+& 2 r_W (2 + r_W + r_W^2), \nonumber\\
w_3^3 &=& 0. 
\label{eq:a.12} 
\end{eqnarray} 

The photon-neutrino scattering formulae  have been obtained assuming non-zero
neutrino mass. The results for massless neutrino agree with the results previously obtained 
(see~\cite{repko} and references therein).
At the end we will complete the notation:
$\alpha$ is electromagnetic coupling, $s$ is square of energy in CM system and 
$\Theta_W$ is Weinberg's Angle.

\section{Appendix B.}
\begin{figure}
\begin{center}
\includegraphics[width=8cm]{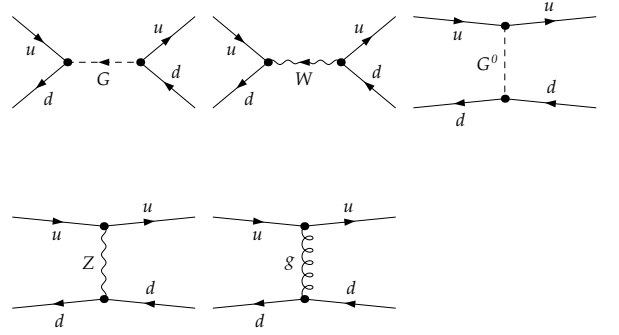}%
\end{center}
\caption{The Feynmann diagrams for the subprocess:  $ud \rightarrow ud$. The contribution in lowest order
of perturbation theory comes from interference between terms represented by $Z^0$/$W$ bosons and gluons
exchanged diagrams.} 
\label{fig:graf}
\end{figure}

To estimate the high energy contributions to p.v. sum rule~(\ref{eq:1.5}) the QCD improved parton model
is used. Polarized parton distributions for proton and unpolarized photon structure have been 
taken from~\cite{grv}.  
In perturbative QCD calculations a lot of subprocesses with quarks and antiquarks from photon and proton participate
 and in principle their contributions might be canceled.
The calculations have been done in the lowest non-vanishing order  in electroweak theory and QCD which gives the 
contribution to p.v. cross sections proportional to $\alpha \alpha_S$. 
Typical example is shown in fig.~\ref{fig:graf} where the diagrams for $ud \rightarrow ud$ quarks interactions 
are drawn.
The p.v. contributions come from the interference terms between subprocesses with $Z^0$/$W$ bosons and gluons exchanged 
between partons.

The total difference of the cross sections is a sum of contributions from all possible quark-quark and quark-antiquark
interactions.
The difference of the cross sections can be expressed by the following formula:

\begin{eqnarray}
\Delta\sigma &=& \frac{32 \pi \alpha }{9}\int \frac{dx_1 dx_2}{x_1 x_2}\frac{d\hat{t}}{\hat{s}^2}
(F_1(x_1,x_2)(g(M_Z^2) + h(M_Z^2))      \nonumber\\
&-& F_2(x_1,x_2) h(M_W^2) - F_3(x_1,x_2) g(M_W^2))
\label{eq:b.1} 
\end{eqnarray}
 where:

\begin{eqnarray}
g(x)&=& \frac{\alpha_S(\hat{t})}{\hat{t}}(\frac{\hat{s}^2}{\hat{u}-x} + \frac{\hat{u}^2}{\hat{s}-x})\\
h(x)&=& \frac{\alpha_S(\hat{s})}{\hat{s}}\frac{\hat{u}^2}{\hat{t}-x}
\label{eq:b.2} 
\end{eqnarray}
and 
\begin{eqnarray}
F_1(x_1,x_2)&=& (x_2 \Delta u(x_2))(x_1 u(x_1))(C_{1u}^2 - C_{2u}^2) \nonumber\\
&+&(x_2 \Delta d(x_2))(x_1 d(x_1))(C_{1d}^2 - C_{2d}^2) \\
F_2(x_1,x_2)&=& C_{3ud}^2((x_2 \Delta u(x_2))(x_1 u(x_1))\nonumber\\
& +& (x_2 \Delta d(x_2))(x_1 d(x_1)))\nonumber\\
&+&C_{3us}^2 (x_2 \Delta u(x_2))(x_1 u(x_1))\\
F_3(x_1,x_2)&=& C_{3ud}^2((x_2 \Delta u(x_2))(x_1 d(x_1))\nonumber\\
& +& (x_2 \Delta d(x_2))(x_1 u(x_1)))\nonumber\\
&+&C_{3us}^2 (x_2 \Delta u(x_2))(x_1 s(x_1))
\label{eq:b.3} 
\end{eqnarray}

The polarized parton distributions in proton are denoted as $\Delta u(x_2)$,$\Delta d(x_2)$ and $\Delta s(x_2)$
for u,d ans s quarks, respectively. The photon structure (unpolarized) is described by $u(x_1), d(x_1)$ and 
$s(x_1)$. The $x_1$ and $x_2$ are the fraction of momentum of photon and proton carried by parton (quark).
The coefficients $C$ are listed below:
\begin{eqnarray}
C_{1u}&=&\frac{2}{3}\frac{sin(\Theta_W)}{cos(\Theta_W)}\nonumber\\
C_{1d}&=&\frac{1}{3}\frac{sin(\Theta_W)}{cos(\Theta_W)}\nonumber\\
C_{2u}&=&\frac{3-4 sin^2(\Theta_W)}{6 sin(\Theta_W)cos(\Theta_W)}\nonumber\\
C_{2d}&=&\frac{3-2 sin^2(\Theta_W)}{6 sin(\Theta_W)cos(\Theta_W)}\nonumber\\
C_{3ud}&=&\frac{CKM_{11}}{\sqrt(2)sin(\Theta_W)}\nonumber\\
C_{3ud}&=&\frac{CKM_{12}}{\sqrt(2)sin(\Theta_W)}
\label{eq:b.4} 
\end{eqnarray}

The integration is performed over kinematically allowed ranges and $\hat{s}=x_1 x_2 s$.
The perturbative scale has been fix as $Q_0^2 = 1 GeV^2$ and $\hat{s} > 4 Q_0^2$.
$CKM_{11}$ and $CKM_{12}$ are Cabbibo-Kobayashi-Maskawa quark-mixing matrix elements.

\begin{figure}
\begin{center}
\includegraphics[width=8cm]{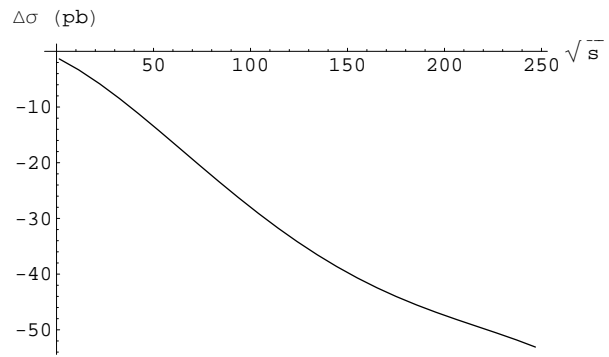}%
\end{center}
\caption{The CMS energy dependence of the difference of the cross sections 
for unpolarized photons scattered off polarized protons, calculated in perturbative QCD.
The parton structure of photon and proton are taken from~\cite{grv}.} 
\label{fig:qcd}
\end{figure}

It is worth to note that the color preservation
forbids some subprocesses to contribute to cross sections. 
The interference terms for right-handed cross sections in the case of different flavours of quarks (e.g. ud,us, etc) give
vanishing contributions when the summation over colors is performed.
The numerical results for the difference of the cross sections as a function of CMS energy 
$\sqrt s$  are shown in fig.~\ref{fig:qcd}.

The differences of the cross sections
are sizeable (up to 60 pb) and rising up (in absolute value sense) as it is shown in 
fig.~\ref{fig:qcd}.
It indicates that for the unpolarized photons case the p.v. sum rule is violated not only 
in the case of point-like lepton targets but also for polarized proton ones. 
The small momenta partons are responsible for the rising of the
absolute value of difference of the cross sections.

\begin{center}
{\bf Acknowledgements}
\end{center}
We would like to thank Professors Norman Dombey and Lech Szymanowski for the discussions.
One of us (L.{\L})was supported in part by the European Community's Human Potential Programme 
under contract HPRN-CT-2002-00311 EURIDICE and by Polish-French Scientific Agreement Polonium.

\end{document}